\documentclass[FBSedit,FBSmath,ecsub]{FBSsuppl}
\usepackage{epsfig}

\def\i{\mathrm{i}}
\def\e{\mathrm{e}}
\def\d{\mathrm{d}}

\def\vec#1{\mbox{\boldmath$#1$}}
\def\svec#1{\mbox{{\scriptsize \boldmath$#1$}}}

\title{Baryons as Solitons in Chiral Quark Models}

\author{B. Golli\instnr{1}\thanks{\textit{E-mail address:} 
bojan.golli@ijs.si}, 
W. Broniowski\instnr{2}, 
G. Ripka\instnr{3}}

\instlist{Faculty of Education, University of Ljubljana and 
J.~Stefan Institute, Jamova 39, P.O.~Box 3000, 1001 Ljubljana, 
Slovenia
\and
H. Niewodnicza\'nski Institute of Nuclear Physics,
         PL-31342 Krak\'ow, Poland
\and
Service de Physique Th\'eorique, Centre d'Etudes de Saclay,
       F-91191 Gif-sur-Yvette Cedex, France, and
ECT${}^\ast$, Villa Tombosi, I-38050 Villazano (Trento), Italy}

\begin{document}

\maketitle
\begin{abstract}
We describe the formation of solitons in NJL-type models
and discuss the influence of the regularization scheme on 
the stability of the solution. We concentrate on models with 
non-local regulators in which stable solutions exist without 
introducing additional constraints.
\end{abstract}

\section{Introduction}

The quark models of baryons can be divided into two main classes:
(i)  {\em constituent quark models\/} 
describing baryons as bound states of three massive quarks 
interacting via a phenomenological potential and
(ii) {\em chiral models\/}
with ``bare'' quarks surrounded by a cloud of either effective 
meson fields or $q\bar{q}$ pairs.
The chiral model distinguishes itself from the constituent 
quark models in that the baryon masses, as well as the 
constituent quark masses are generated dynamically.
The two classes of models describe hadrons at two different
levels: at a higher level,
the constituent quark model successfully predicts the spectrum 
of baryons and their excited states; 
at a lower level, the chiral models are able to describe 
the vacuum, meson and baryon sectors with the same effective 
Lagrangian and they model the origin of the masses of constituent 
quarks and in particular the role the chiral mesons play in 
the constituent quark model.

We present here the calculation of solitons in NJL-type models 
and discuss how different regularization schemes influence
the stability of solutions. We focus on the version with non-local 
quark interactions as suggested from the instanton-liquid model
which supports stable solutions without introducing artificial 
constraints as in previous calculations
\cite{nls,solilong}.

\section{Basic Properties of the NJL}

The NJL model assumes an attractive interaction between quarks 
constructed in terms of quark bilinears in a form that obeys the 
chiral symmetry.
To describe low energy phenomena, it is enough to consider the 
non-strange quarks and keep only the scalar and pseudoscalar terms. 
(For a review of the model and its applications see 
ref.~\cite{NJLrev}.) In the chiral limit, the  Lagrangian becomes:
\begin{equation}
   \mathcal{L}_{NJL} = 
     \bar{q}\left(\i\partial_\mu\gamma^\mu\right)q
  + {G\over2}\left[(\bar{q}q)^2 
          + (\bar{q}\,\i\gamma_5 \tau_a q)^2\right]\;,
\label{L}
\end{equation}
where $a = 1,\ldots 3$ is the isospin index.
The interaction is point-like and the UV divergences have 
to be removed by a suitable regularization prescription.

To solve the model in the mean-field (Hartree) approximation
it is practical to introduce auxiliary chiral fields 
\begin{equation}
          S = -G\; \bar{q}q
\qquad
\mbox{and}
\qquad
          P_a = -G\;  \bar{q}\,\i\gamma_5\tau_a q\;,
\label{fields}
\end{equation}
such that the Lagrangian takes the {\em semi-bosonized\/} form:
\begin{equation}
   \mathcal{L}_{NJL} = 
     \bar{q}\beta\left(\i\partial_t - h \right)q
  - {1\over 2G}\left(S^2 + P_a^2\right)\;,
\label{Lboson}
\end{equation}
\begin{equation}
     h = -\i\vec{\alpha}\cdot\vec{\nabla}
         + \beta (S + \i\gamma_5\tau_a P_a)\;.
\label{h}
\end{equation}

The vacuum in the Hartree approximation is calculated by assuming 
a constant value, $M$, for $\langle S\rangle$ and evaluating 
$\langle\bar{q}q \rangle$ in (\ref{fields}) by summing up the bubble 
graphs or, equivalently, by performing functional integration over 
the quark fields. This yields the so called ``gap equation'':
\begin{equation}
       {1\over G} = 24 \sum_\Lambda{1\over k^2 + {M}^2}\;,
\end{equation}
where $\sum_\Lambda$ denotes a regularized sum over quark states 
with a cut-off parameter $\Lambda$. Starting from a zero mass we 
have generated a finite quark mass. The model thus exhibits the 
{\em spontaneous breaking of the chiral symmetry}, yielding a  
nonzero value for the vacuum quark condensate
$\langle\bar{q}q \rangle = -M/G$, and three Goldstone bosons (pions) 
corresponding to small oscillation of the pseudoscalar field around 
its vacuum value $\langle P_a \rangle = 0$.
The model predicts the pion decay constant, $f_\pi$, and for a finite 
current quark mass, $m\ne 0$, also a finite pion mass satisfying 
the Gell-Mann--Oakes--Renner relation. 
Two of the three free parameters, $G$, $\Lambda$, 
or $m$ can be eliminated by fitting $f_\pi$ and $m_\pi$.
The remaining free parameter is usually expressed in term of  $M$,
the so-called {\em  constituent mass\/}.

In calculations of mesons and baryons several regularization 
schemes have been used including a {\em sharp cut-off\/} in 3- 
or 4-dimensions, the {\em proper-time\/} and {\em Pauli-Villars\/} 
regularizations and {\em non-local regulators\/} as suggested 
from the instanton-liquid model. All of them require a rather 
low cut-off parameter, $\Lambda \sim 600$~MeV, which indicates 
that it may be derivable from the underlying theory. 
Indeed, the instanton-liquid model \cite{Shuryak} predicts 
a non-local interaction between quarks with $\Lambda$ related 
to the inter-instanton spacing $\rho^{-1} \sim 600$~MeV.  
The calculated vacuum properties (quark condensate, gluon 
condensate) suggest the value $M$ should lie in the range 
between 300 and 400~MeV.

The soliton, corresponding to a baryon, is constructed around 
three valence quarks that polarize the vacuum generating a 
spatially non-trivial configuration. 
The resulting mean-field potential may make the 
system stable against decaying into free quarks.
This is not a mechanism of confinement but simply a mechanism 
to bind quarks. If this scenario is realized it means that 
the interactions of the quarks with the chiral field may play 
an equally important role in the formation of the baryon as 
the color confining forces.

The existence of a stable solution is closely related to the 
choice of regularization scheme. A simple sharp cut-off does not 
yield stable solutions. Using the proper-time or Pauli-Villars 
regularization a stable soliton can be formed only by restricting 
the chiral fields to lie on the chiral circle, i.e.
$S(\vec{r})^2 + P_a(\vec{r})^2 = M^2$, otherwise the scalar field
acquires an arbitrarily low value at the origin producing a strongly 
localized state with a zero energy. Such a constraint does not have 
any justification in the model. It is only in the version with 
non-local regulators that solitons exist without introducing 
additional constraints.

\section{Solitons in the NJL Model with Non-Local Regulators}

In a non-local version of the NJL, the $q\bar{q}$ interaction 
is smeared by replacing the quark fields in the interaction part 
of (\ref{L}) by {\em delocalized quark fields\/}:
\begin{equation}
  \psi(x) = \langle x|\psi\rangle = \langle x|r|q\rangle
          = \sum_k r_k e^{\i kx}\langle k|q\rangle \;.
\end{equation}
Here $r_k$ is a regulator, diagonal  in  $k$-space, and $k$ is 
the {\em Euclidean} 4-momentum, $k^2 = \omega^2 + \vec{k}^2$.
The instanton-liquid model predicts a form of the regulator.
For model calculations it is simpler to take a Gaussian form,
$
        r_k=e^{-k^2/2\Lambda^2}
$.
Using the Euclidean formulation the Dirac Hamiltonian takes the 
form
\begin{equation}
     h = -\i\vec{\alpha}\cdot\vec{\nabla}
         + r\beta Sr + \i\gamma_5\beta\tau_a rP_a r\;,
\qquad
      r = r\left(-\partial_t^2-\vec{\nabla}^2,\Lambda\right)\;.
\end{equation}

The key point is the construction of the valence orbit.
Let us first study the free quark propagator. 
In the chiral limit  ($m=0$) the inverse propagator takes 
the form $k_\mu\gamma_\mu + r(k^2)^2M$.
The pole (setting $\vec{k}^2=0$, $k_0^2=-M_q^2$) occurs at the 
solution of
\begin{equation}
        {M_q}^2 = M^2 \e^{2{M_q}^2/\Lambda^2}\;.
\label{Mqeq} 
\end{equation}
The most striking feature of (\ref{Mqeq}) is that the solution
exists only 
below a critical value of $M$. Above this value, free on-shell 
quarks do not exist in the vacuum.

The treatment of the quark propagator in a spatially non-homogeneous 
field configuration of the soliton is a non-trivial problem because of 
the presence of the time-dependent regulator in the Dirac Hamiltonian.
For stationary chiral fields $S$ and $P_a$ the Dirac operator $h$ is 
diagonal in the energy representation:
\begin{equation}
    h(\omega^2)\,|\,\lambda_\omega\,\rangle = 
          \i\omega\,|\,\lambda_\omega\,\rangle\;.
\label{Diraceq}
\end{equation}
The lowest (valence) orbit is obtained as a solution of 
Eq. (\ref{Diraceq}) for  $\i\omega \rightarrow {\varepsilon_0}$:
\begin{equation}
\left[{\vec{\alpha}\cdot\vec{\nabla}\over\i}
         + \beta\,
      \e^{{1\over2\Lambda^2}\,({\varepsilon_0}^2+\svec{\nabla}^2)}
           \left({S(\vec{r})} + \i\gamma_5\tau_a {P_a(\vec{r})}\right)
      \e^{{1\over2\Lambda^2}\,({\varepsilon_0}^2+\svec{\nabla}^2)}\right]
|\lambda_0\rangle = {\varepsilon_0}  |\lambda_0\rangle \;. 
\label{Diracval}
\end{equation}
The solution of Eq.~(\ref{Diracval}) exists for all $M$ beyond 
$M_\mathrm{cr}$ as shown in Fig.~\ref{fige}.

The soliton is sought iteratively by assuming a hedgehog shape for 
the chiral fields. Starting from an initial guess for 
$S(\vec{r})$ and $P_a(\vec{r})$ Eq.~(\ref{Diraceq}) is solved
for the valence orbit, $|\lambda_0\rangle$, and the sea orbits, 
$|\lambda_\omega\rangle$. In the next iterations new values for the 
chiral fields are obtained using the Euler-Lagrange equations:
\begin{eqnarray}
\frac{S( \vec{r}) }{G} &=&N_{c}{\rm z}_0\langle 
\lambda_0|r|\vec{r}\rangle \beta \langle \vec{r}|r|\lambda_0\rangle
+\int
\frac{d\omega }{2\pi }\sum_{\lambda _{\omega }}%
\frac{\langle \lambda _{\omega }|r|\vec{r}\rangle \beta \langle \vec{r}%
|r|\lambda _{\omega }\rangle }{\i\omega +e_{\lambda }\left( \omega
^{2}\right) },  
\nonumber \\
\frac{P_{a}( \vec{r}) }{G} &=&N_{c}{\rm z}_0\langle 
\lambda_0|r|\vec{r}\rangle \i\beta \gamma _{5}\tau _{a}
\langle \vec{r}|r|\lambda_0\rangle +\int
\frac{d\omega }{2\pi }\sum_{\lambda
_{\omega }}\frac{\langle \lambda _{\omega }|r|\vec{r}\rangle \i\beta \gamma
_{5}\tau _{a}\langle \vec{r}|r|\lambda _{\omega }\rangle }{\i\omega
+e_{\lambda }\left( \omega ^{2}\right) }.  
\nonumber
\end{eqnarray}
Here $z_0$ is the residue factor:
${\rm z}_0=\left( 1-\i\left. \d \varepsilon_0(\omega )/\d\omega
\right|_{\omega =\i\varepsilon_0}\right)^{-1}$.
The convergence is obtained for all values of $M$ beyond
a critical value $M_\mathrm{cr}$.

In this approach the soliton acquires the correct baryon number 
\cite{nls}.

\section{Properties of the Soliton}

\begin{figure}[b]
\vspace{0mm} \centerline{\includegraphics[width=6.3cm]{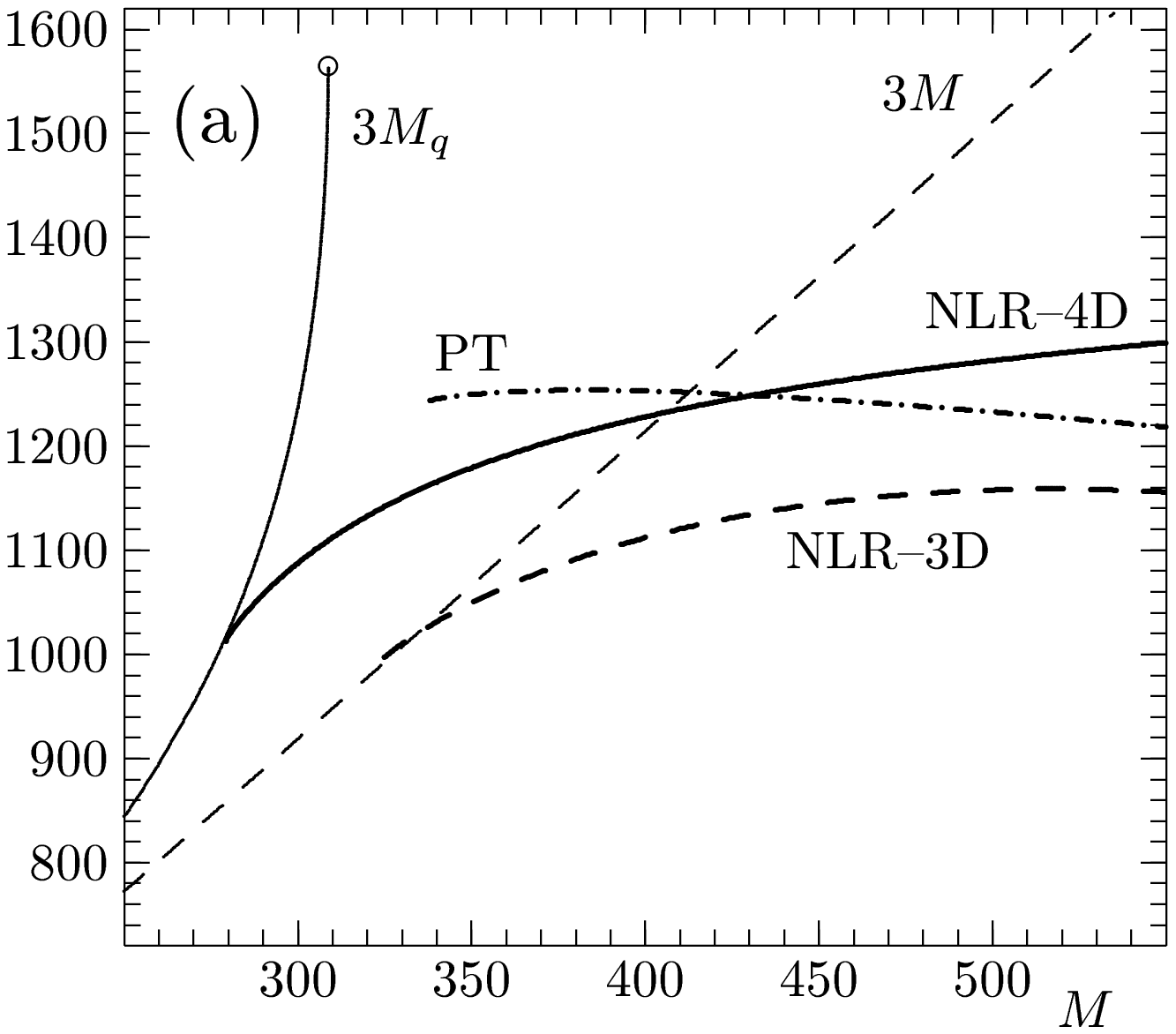}
\kern3mm\includegraphics[width=5.4cm]{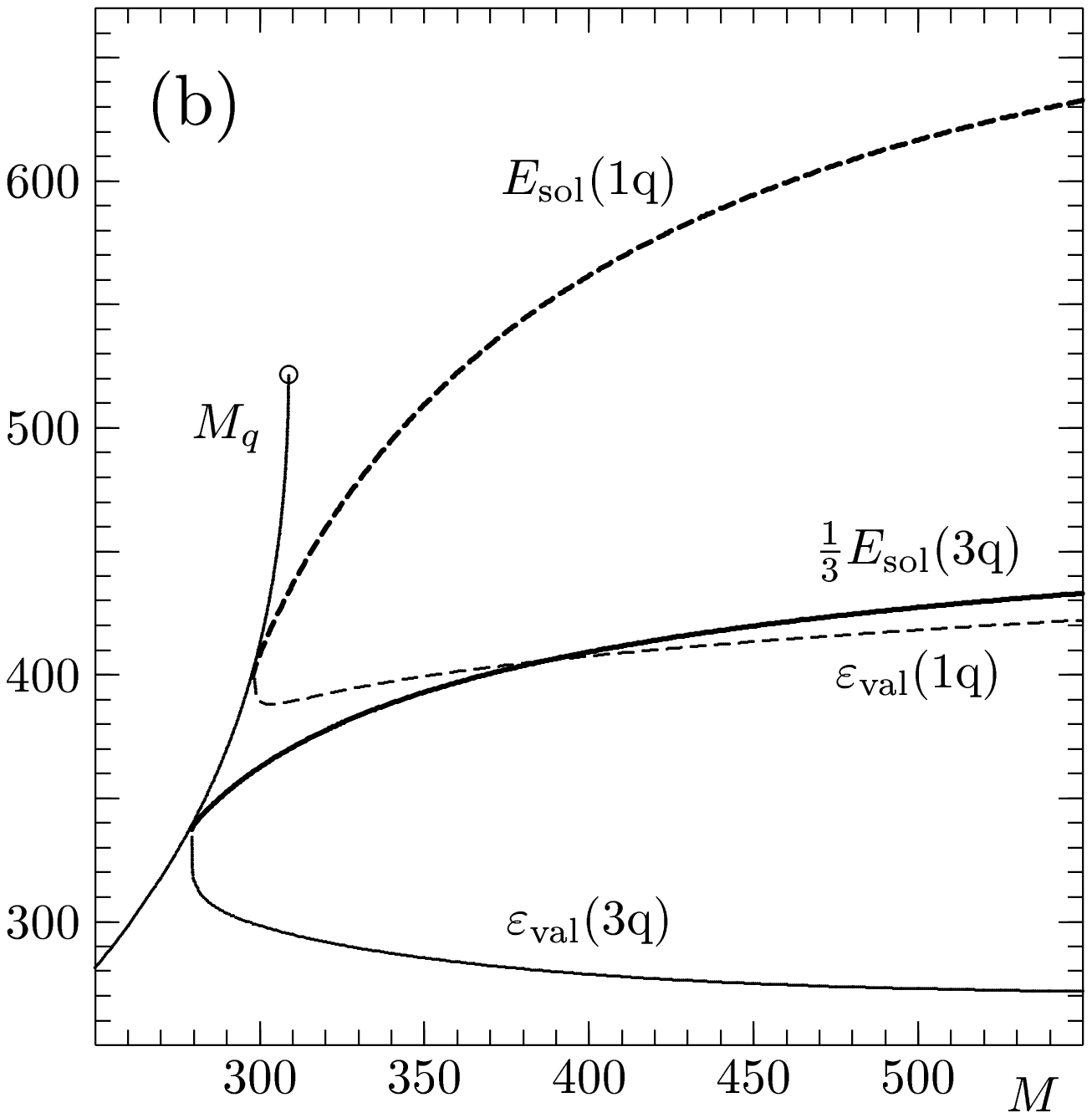}} 
\vspace{0mm} 
\caption{(a) The energy of the soliton for different regularization
schemes plotted as functions of the parameter $M$ using the 
4-dimensional non-local Gaussian regulator (solid line), 
the 3-dimensional Gaussian regulator (dashed line), and the 
proper-time regularization (dashed-dotted line).
(b) The energy per quark of the soliton with three valence 
quarks (solid line) and with  one valence quark (dashed line)
plotted as functions of the parameter $M$. 
Also shown are the corresponding valence energies. 
$M_q$ is the free-space on-shell quark mass.
All quantities in MeV. For each case the values of $f_\pi$ and 
$m_\pi$ have been fixed to their physical values.}
\label{fige}
\end{figure}

Figure~\ref{fige} (a) shows the soliton energy as a function of
$M$ for three regularization schemes: 
(i) the non-local model described in Sect.~3 yields stable solutions 
above $M_\mathrm{cr}=280$~MeV which are also energetically stable 
since the energy of three  free on-shell quarks is always higher 
than the soliton energy,  
(ii) the model with a non-local regulator of the Gaussian shape
that depends only on the 3-momentum $\vec{k}$ possesses stable 
solutions above $M\approx325$~MeV; however, beyond $M\approx600$~MeV 
the solution is again unstable since it becomes energetically 
favorable for two quarks in the soliton to acquire high momenta,
lower their effective mass, and escape from the third quark
(Such an awkward behavior is a consequence of breaking the Lorentz
invariance and does not show up in the model with the 4-momentum 
regulator.) \cite{bled99},
(iii) the model using the proper time regularization and 
the chiral-circle constraint has stable solutions above $M=340$~MeV 
that are energetically stable only beyond $M=400$~MeV.

The model with the 4-momentum non-local regulator supports another 
type of solutions with one or two quarks polarizing the sea 
(Fig.~\ref{fige}).
The solution with three  valence quarks  is stable against
disintegrating into solitons with a lower number of valence quarks.
This does not hold for a soliton with four or more valence quarks 
since in such a case the additional quarks would have to fill 
the grand spin 1 orbit which is energetically less favorable.

\begin{figure}[tb]
\vspace{0mm}  \centerline{\includegraphics[height=45mm]{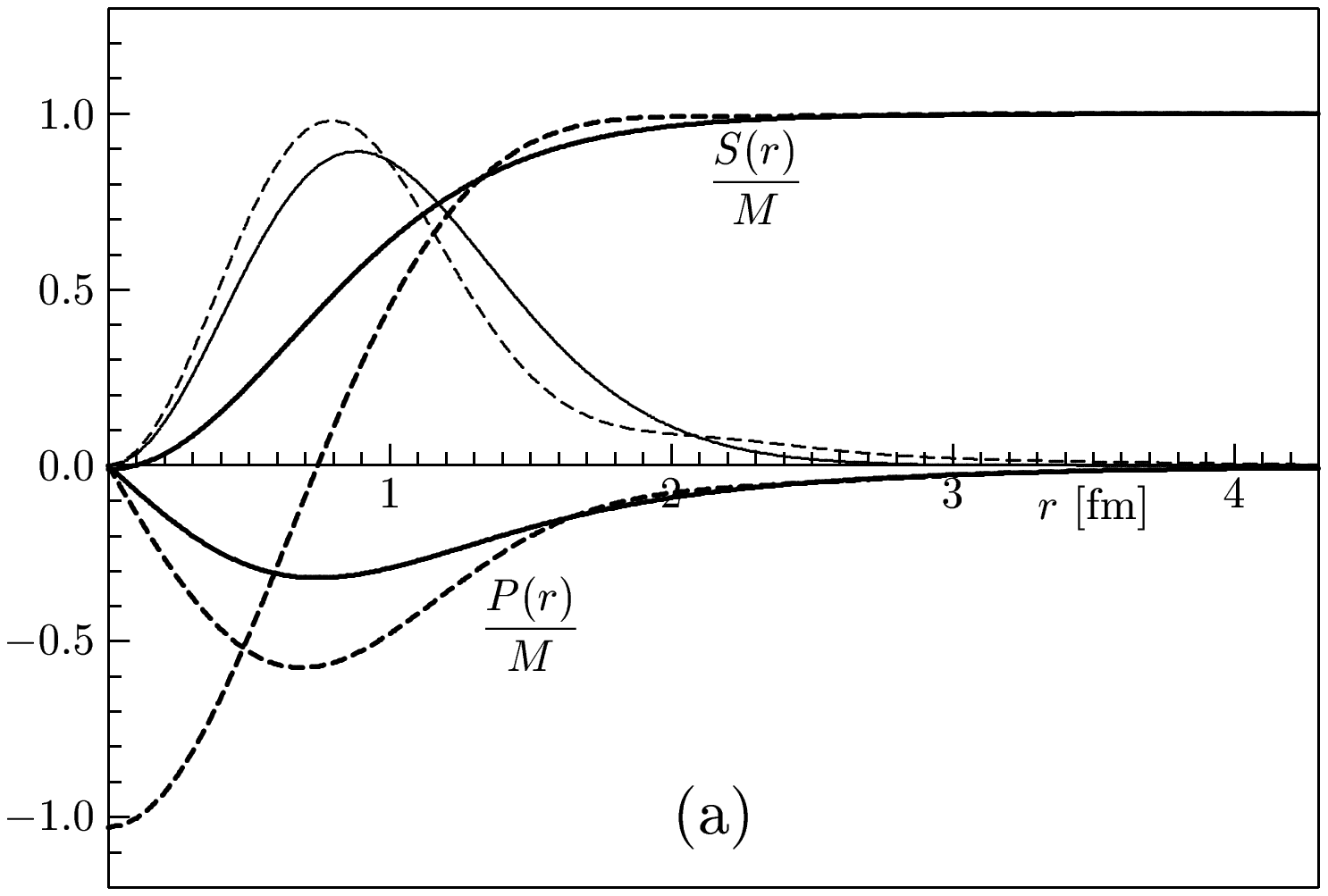}
\kern3mm\includegraphics[height=42mm]{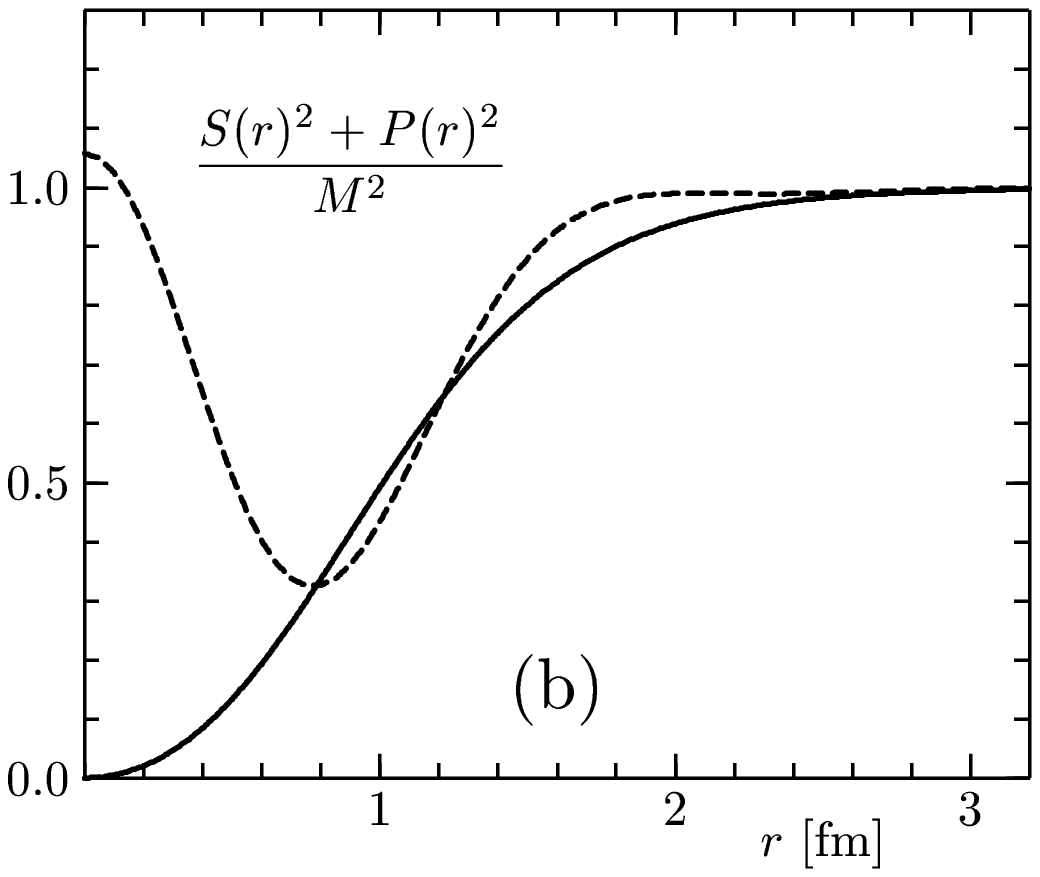}} 
\vspace{0mm} 
\caption{(a) Self consistently determined fields 
and baryon densities (multiplied by $4\pi r^2$);
(b)  effective squared quark mass for $M=325$~MeV (solid line) and 
$M=750$~MeV (dashed line), plotted as functions of the radial 
coordinate $r$.}
\label{figmf}
\end{figure}

The calculation of observables in the model with non-local 
regulator is a non-trivial task since the Noether currents 
acquire additional terms originating from the momentum dependent 
regulator~\cite{solilong}.
Fortunately, their contribution remains small for physically 
interesting values of $M$ below 400~MeV.
The calculated properties reflect a relatively large size of 
the soliton compared to solutions using e.g. the proper time 
regularization as well as to the experimental values.
In our opinion this is a consequence of the mean-field treatment 
rather than a serious deficiency of the model.
Any improvement of the approximation such as elimination of 
spurious center-of-mass motion or projection onto subspace with 
good spin and isospin may considerably reduce the size \cite{coim}.

\section{Implications for Other Models}

The solitons constructed in the chiral model with non-local
regulators share several features found in more phenomenological 
models. From the model it is possible to derive a version of 
the linear $\sigma$-model that approximates well the full 
model \cite{njlsigma}.

The effective quark mass (i.e the square root of $S^2+P^2$ 
shown in Fig.~\ref{figmf}) drops almost to 0 in the center of 
the soliton for $M$ between 300~MeV and 350~MeV similarly as 
in models based on restoration of the chiral symmetry inside
the baryon like different versions of the bag model.

The Goldstone-boson exchange constituent quark model also 
finds a qualitative support in the non-local model.
The solitons consisting of only one valence quark (see 
Fig.~\ref{fige}) can be interpreted as constituent quarks. 
Above $M\approx310$~MeV these solutions are dressed in 
a cloud of chiral mesons that generate attraction and a 
3-quark soliton is formed.
The interaction between such objects calculated in the linear 
$\sigma$-model \cite{GR97} shows a typical behavior of the 
potential used in the Goldstone-meson exchange constituent 
quark model.

\bigskip\noindent
{\em Acknowledgment.\/} This work was supported in part by the
Scientific and Technological Cooperation Joint Project between
Poland and Slovenia, financed by the Ministry of Science of 
Slovenia and the Polish State Committee for Scientific Research, 
and by the Polish State Committee for Scientific Research, grant 
number 2 P03 09419.

\end{document}